# A strategic framework for identifying the critical factors of 4G technology diffusion in I.R. Iran - A Fuzzy DEMATEL approach

Hossein Sabzian[1], Hossein Gharib[2], Seyyed Mostafa Seyyed Hashemi[3], Ali Maleki[4]

1: Iran University of Science and Technology, Tehran, Iran.
2: Tadbir Economic Development Group, Tehran, Iran.
3: Tadbir Economic Development Group, Tehran, Iran.
4: Sharif University of Technology, Tehran, Iran.

ABSTRACT

As the most prominent representative of 4G, Long term evolution (LTE) technology has become a focal point for mobile network operators all over the world. However, although Iranian main operators like MCI and Irancell have hugely invested on deployment of this technology, its diffusion has been very slow with a penetration rate of 0.06 at the end of spring 2017. Nevertheless, if this rate doesn't increase, it will yield some negative unintended consequences for telecom operators such as (I) Failure to provide a large number of high quality services (II) Inability to compete with OTT technologies (III) Loss of many revenue opportunities (IV) Prolongation of payback period and (V) The lack of technological integrability with fifth generation networks (5G) and loss of many IOT opportunities. Through discussing the literature of technology adoption and diffusion both generally and specifically, identifying the major limitations of these studies and establishing a comprehensive factor set based on four major groups of (I) mobile handset and operators-related factors (II) subscribers-related biological factors, (III) subscribers-related perceptual factors and (IV) subscribers-related contextual factors, a novel fuzzy DEMATEL model has been developed by which all ICT policy makers can not only get a clear knowledge of factors influencing technology adoption but also know the critical success factors (CSFs) influencing Iranians' mindsets towards LTE adoption. Therefore, they can make effective and actionable policies to scale up LTE diffusion or other ICT-related technologies throughout the society.

*Keywords; Technology adoption and diffusion, Fourth generation technology (4G), Long term evolution (LTE), Critical success factors (CSFs), Fuzzy Logic, DEMATE*

I. INTRODUCTION

High-speed access to mobile data has become one of the most important demands of smart phone subscribers. The increasing availability of intelligent equipment and ever-growing demand for multi-media streaming services have considerably increased the volume of mobile data traffic [10]. In 2016, the total generated mobile data was 8.8 Exabytes 50% of which belonged to video contents. It is expected that this amount will reach 71 Exabytes in 2020 75% of which will belong to video contents [12]. Network carriers have spent lots of R&D costs on providing better data. Therefore, it has led to significant expansion of mobile broadband. Statistics and figures indicate that mobile broadband subscriber growth rate has been steadily higher than that of fixed broadband in the world. Fixed broadband penetration rate has reached 11.9% in 2016 from 0.9% in 2012 while mobile broadband penetration rate has reached 49.4% in 2016 from 21.7% in 2012 (Figure 1). In I.R. Iran, mobile broadband penetration rate has been much higher than that of fixed broadband as the growth rate of fixed broadband subscribers has moved from 2.81% in the summer 2012 only to 12% in the spring 2017 while the growth rate of mobile broadband subscribers has shifted from 1.92% in summer 2012 to 42.5% in spring 2017 (Figure 2).

Cellular technologies such as the third generation (3G) and fourth generation (4G) of mobile networks have majorly contributed to the development of mobile broadband. 3G Technology was developed during the 90's and entered global market in 2002. This technology was getting widespread in 2007. Actually, in December 2007, 190 operators in 40 countries provided 3G services to their customers. In I.R. Iran, Rightel operator which was established in 2007, received the 3G service license in January 2010 but the operator started its work in June 2011. In I.R. Iran, 3G mobile networks entered the market with a delay of 9 years. In fact, the commercialization of 3G in I.R. Iran coincided with the commercialization of 4G in the world. In I.R. Iran, the fourth generation of mobile networks was launched in September 2014 by the I.R. Irancell operator. Subsequently, mobile operators such as Mobile Telecommunication Company of Iran (MCI) and Rightel also launched 4G mobile networks. Limitations such as low-rate of transmission, mobility and bandwidth constraints were among the most important reasons leading mobile operators to go to launch 4G mobile networks.

---

[1] : Corresponding author: hossein_sabzian@pgre.iust.ac.ir

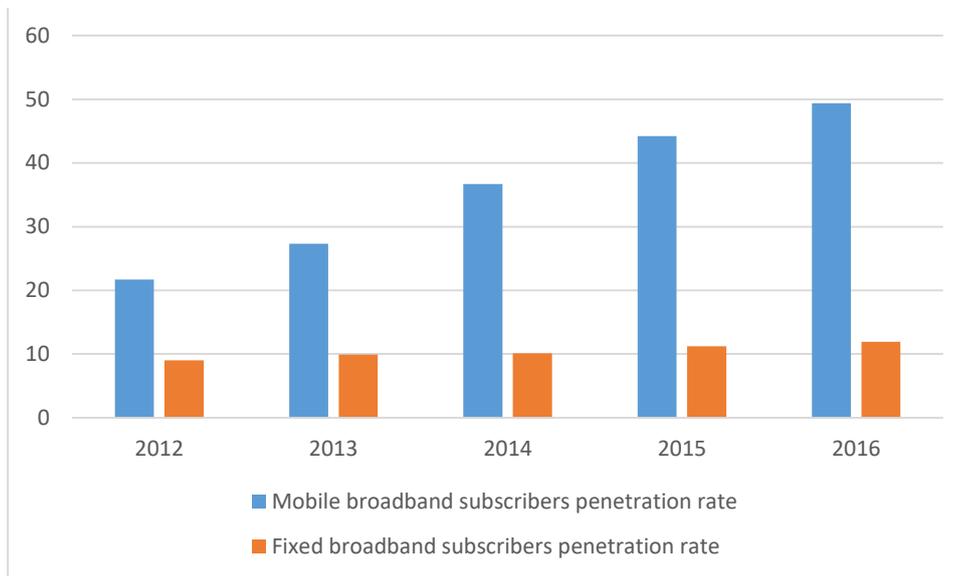

*Figure 1: Comparison of penetration rate of mobile broadband subscribers with that of fixed broadband subscribers in the world from 2012 – 2016 ( Source: ITU [67])*

…

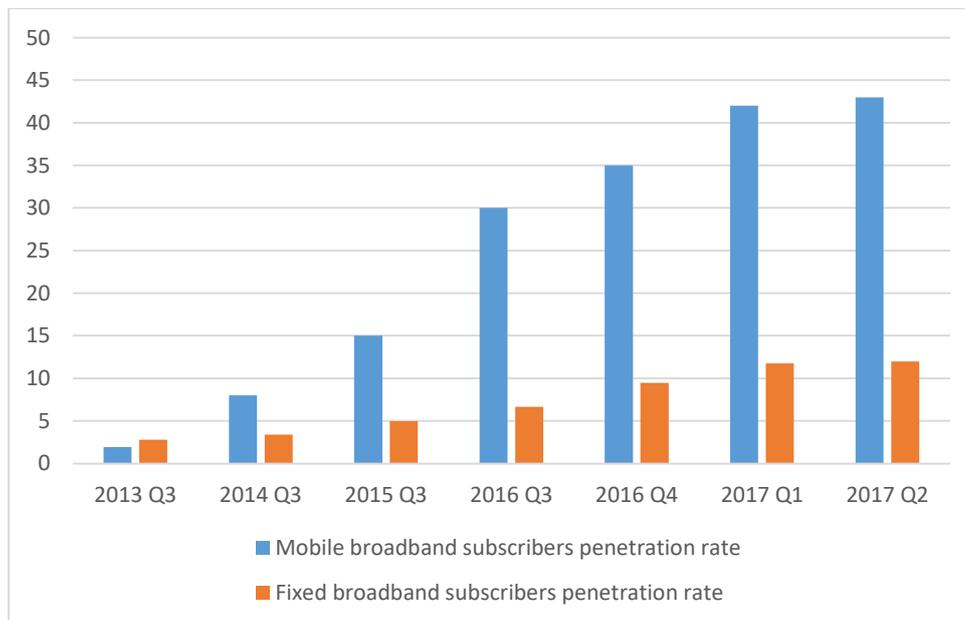

*Figure 2: Comparison of penetration rate of mobile broadband subscribers with that of fixed broadband subscribers in the I.R. Iran from 2013 Q3 – 2017 Q2 ( Source: Ministry of I.C.T of I.R. Iran)*

According to specifications detailed by IMT-advanced, any 4G mobile network should meet the following minimum requirements:

- It should be completely based on Internet Protocol (IPV6)
- Data transmission rate should be 100 megabits per second (Mbit/s) for high mobility communications (such as from trains and cars) and 1 gigabit per second (Gbits) for low mobility communications (such as pedestrians and stationary users)

As the most prominent representative of 4G, Long Term Evolution (LTE) technology has become very popular. LTE was developed in 2004 and commercialized during 2010-2011. LTE which is often marketed as 4GLTE is a standard for transmitting high-speed wireless data for data terminals and mobile devices. LTE is based on GSM/EDGE and UMTS/HSPA technologies that has greatly increased the capacity and speed using a different radio interface together with core network improvements. LTE standard has been developed by 3$^{rd}$ Generation Partnership Project. Differences among mobile network have been presented in Table 1

Several studies have indicated the widespread market share of LTE in the next five years so that if mobile operators cannot properly manage this technology, they will eventually lose much of their subscriber's market share. The Ericsson 2016 report has illustrated this trend (Figure 3). According to this report, there were 3.9 billion smartphone subscribers in 2016, of which 1700 million are 4G, 1800 million are 3G (UMTS/HSPA), 300 million are 2G (GSM/EDGE) and another one million subscribers use other cellular mobile networks. It is expected that by the year 2022, 2700 million persons will be added to 4G subscribers and 540 million persons to 5G subscribers while 3G, 2G and other mobile network technologies will lose 65 million, 95 million and 130 million subscribers respectively.

*Table 1: Different generations of mobile technologies from 1G to 4G (adapted from [73].)*

| Item | 1G | 2G | 3G | 4G |
|---|---|---|---|---|
| Development time | 1970 | 1982 | 1990 | 2004 |
| Commercialization time | 1984 | 1991 | 2002 | 2010-2011 |
| Technical features | Analog voice | Digital voice Short messages | Live podcast speed up to 2Mbps, vide form of telecommunication | IP-based, Multimedia, Long distance transmission |
| International standard | AMPS, TACS, NMT | GSM, TDMA, DAMPS, CDMA, PDC | WCDMA, CDMA2000, DETC, TDSCDMA | 3GPP |
| Speed | 1.9 Kbps | 9.6-172 Kbps | 382 Kbps-2 Mbps | 86-326 Mbps |
| Application | Voice | Voice and Data | Voice, Data, video call | Voice, Data, video call, Online game, HD TV |

Since the introduction of LTE technology for the first time in I.R. Iran (September 2014), Iranian operators have been investing heavily in developing this technology. The increase in coverage of this network from 47.86% in November 2016 to 58.82% in June 2017 indicates an increase in the number of 4G transmitter / receiver base stations (BTS) across I.R. Iran. In addition, the share of mobile data demand is increasing day by day in I.R. Iran, which is considered as one of the strongest drivers for average revenue per user (ARPU).

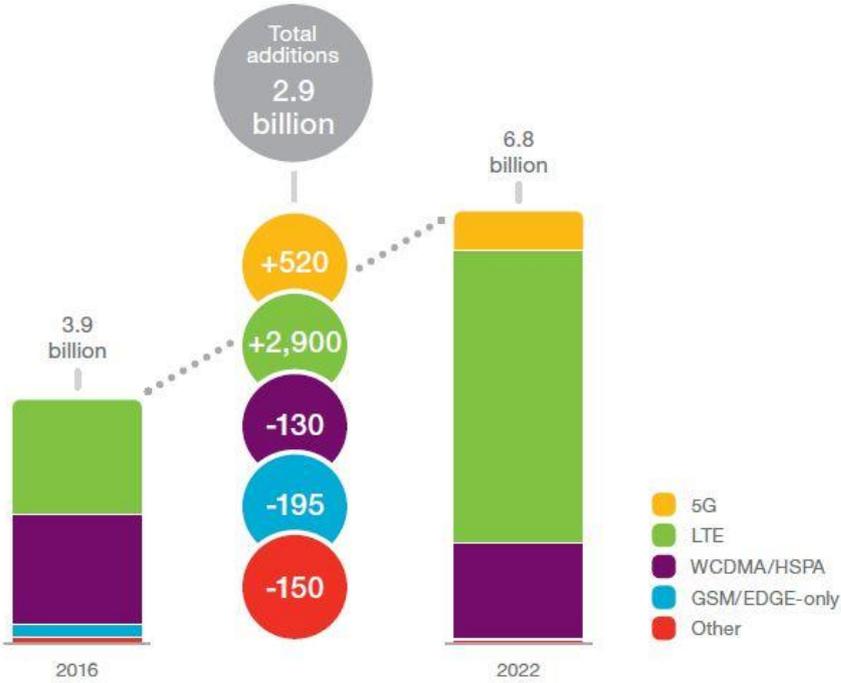

*Figure 3: Number of smart phone users according to type of cellular technology ( Source: [12])*

According to an analysis by Ministry of Information and Communications Technology (ICT) of I.R Iran, the total portfolio of Iranian operators in the year 2015 was 20,000 trillion tomans, 74.1% of which was voice and SMS and only 4.6% was mobile data. With the development of mobile infrastructure and the availability of high-speed broadband, as well as the widespread diffusion of various OTT technologies such as Telegram a penetration rate of 78%, and Instagram with a penetration rate of 54% by the end of spring 2017, it is forecast that of the 67,000 billion tomans revenues for Iranian operators in 2020, only 35.1% would be voice and SMS and 26.9% for mobile data. It reflects the fact that by 2022 the share of voice and SMS will be halved and the data share will roughly get six times higher. The portfolio revenue change of Iranian operators is shown in Figure 4.

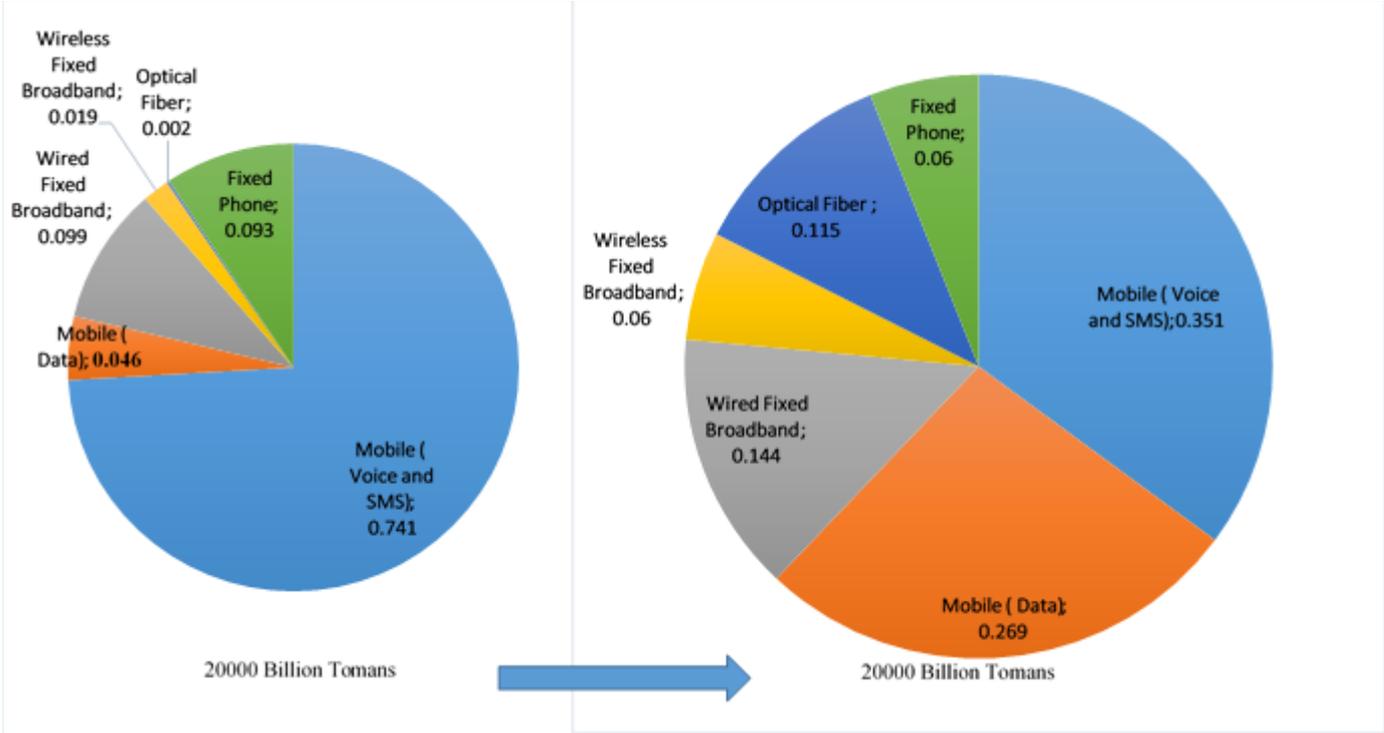

*Figure 4 : Evolution of revenue portfolio of I.R. Iran's Operators from 2015 – 2020 ( Source：Ministry of I.C.T of I.R. Iran)*

However, in spite of the large investments of the three main operators of I.R. Iran, namely MCI, Irancell and Rightel, in the development of 4G LTE network, as well as the increasing penetration of mobile broadband in I.R. Iran (about 42.5% by the end of the spring of 2017), The LTE mobile penetration rate in I.R. Iran is estimated at 6% by the end of spring 2017. This fact indicates that the main volume of mobile broadband subscribers in I.R. Iran is still using UMTS / HSPA 3G networks. In other words, about 29 million of total mobile broadband subscribers (i.e., 34 million subscribers by the end of spring 2017) are currently using 3G cellular networks. Given the mobile broadband penetration rate in I.R. Iran and the third-generation share in it (85％), it is clear that mobile broadband subscribers lock on these third-generation UMTS / HSPA networks and the poor reception of the fourth generation will slow down future investments in LTE networks. This can result in some negative unintended consequences as following:

- **Failure to provide a large number of high quality services:** With the help of LTE technology, operators can provide much more diverse and high-quality services. Some of the major advantages of this technology include faster mobile data, more bandwidth, much less latency, better management of data traffic, high-level video streaming, Voice over LTE (VoLTE), Video over LTE (ViLTE), entry to vertical markets (e.g., public safety, health and transportation), low legacy cost due to more integrability with fifth generation networks (5G), network better efficiency, cell broadcast and lots of value-added services

- **Inability to compete with OTT technologies:** Because LTE operators can provide better quality services than OTT technologies, such as VoLTE versus Voice over Internet Protocol (VOIP) offered by whatsapp or ViLTE compared to the video over the Internet protocol provided by Skype

- **Loss of many revenue opportunities**: The revenue generated by LTE is forecast to be $ 350 billion by 2020 [69]. Considering that I.R. Iran is 1.08% of the world's population, I.R. Iran's share of this amount can be equal to 3.780 billion. Therefore, if the fourth generation is not well understood in I.R. Iran, Iranian operators will lose that income share.

- **Prolongation of payback period:** If operators' subscribers migrate slowly from 3G to 4G, this can prolong the payback period of 4G network and create a lot of costs for operators

- **The lack of integrability with fifth generation networks (5G) and loss of many IOT opportunities**: IOT is recognized to have a tremendous effect over ICT industry in the near future. The Forbes Media Company predicts that the IOT revenue share will be $ 14.4 trillion by 2022. Of this, $ 2.5 trillion has been spent on improving employees' productivity, another 2.5 trillion in reducing costs, 2.7 trillion in improving logistics and supply chains, 3 trillion in reducing market entry time, and 3.7 trillion in improving the customers' experience [56]. 5G plays a pivotal role in realizing the IOT and utilizing its capacities. 4G technologies, in particular LTE-Advanced Pro technology, have a high potential to integrate with 5G in comparison to other cellular technologies. Via this technology, operators can get integrated with 5G networks without getting stuck at a huge legacy cost. Therefore, if the operators only stay on 3G networks, huge costs will be imposed on them to migrate to 5G Networks, and moreover, they will lose huge opportunities.

Therefore, it is important for managers of Iranian operators to be aware of how they can encourage third-generation subscribers to migrate to fourth-generation technologies. The purpose of this study is to provide such knowledge. In this study, by reviewing a large number of theories and technology diffusion models, a set of critical factors that have the greatest impact on the fourth-generation technology of mobile networks in the Iranian community has been identified. Being informed of such critical factors enables domestic operators' strategists to formulate effective strategies for 4G diffusion in I.R. Iran market.

A number of studies have been conducted on identification of critical factors driving adoption of mobile services. In terms of methodology, such studies can be classified into two classes of 1) statistical regression-based studies [1,2,4,5,38] and 2) Multi-Criteria Decision Making (MCDM) ones [48,49,50,51,52,53]. Statistical regression based models have some basic limitations. The first limitation is that such models are formulated based on some simplistic assumptions which can distort the accuracy of solutions. For example, in linear regression, the variables are considered to be linearly related om which the independent variables can take a range of values while dependent ones have a random value. In addition, successive observations of independent variables are regarded to be uncorrelated and the co-linearity among variables should be taken into consideration. The second limitation is that the prediction error is very likely in such models, therefore, there can be a great gap between the outcome variable and real-world data. The third limitation is that in multiple regression, the beta value is indirectly calculated by a test so in case of a negative beta, the final judgment becomes really onerous. MCDM studies have been frequently used to pinpoint the major criteria affecting adoption of mobile services. Liu (2009) used Analytical Hierarchy Process (AHP) to rank the critical factors of mobile handset usage according to employees' opinions. In that study, the critical factors were divided into four factors of (1) economic value, (2) relational value, (3) knowledge value and (4) convenience value that the convenience value was finally selected as the most critical factor [48]. Phan and Diam (2011) ranked and clustered the major influencers on mobile service adoption using AHP and cluster analysis. Their study included 5 total factors of habits, social factors, technology, ease of use and usefulness. The results showed that the last 2 factors play a vital role in adoption process [49]. Regardless of business aspects, Nikou and Mezei (2013) could rank the factors impacting students to adopt mobile services using an AHP approach. They identified 20 mobile services in five distinct categories. The results of their study indicated that five services of "short message"," mobile email", "mobile internet surfing", "mobile search" and "mobile Google map" were believed to be the most important services affecting mobile adoption [50]. However, none of these studies dealt with the diffusion problem in a crisp manner So the lack of a fuzzy assessment is their biggest limitation. Buyukozkan (2009) developed a fuzzy AHP (FAHP) to prioritize requirements of mobile experiences according to organizational criteria and opinions expressed by users and experts. However, considering a small number of criteria is considered to be the prime limitation of this study [52]. Combining FAHP and extent analysis approach, Lin (2013) developed a fuzzy assessment model for prioritizing factors affecting the quality of mobile banking. The opinions of two different groups were used in this study one of which possessed a little experience of mobile banking while another one had a high level of experience in it. However, the results show that two groups had chosen the criterion of "customer services" as the most critical factor influencing the effectiveness of electronic banking [53]. Sheih and colleagues (2014) used FAHP to rank critical factors of adoption of mobile services by Taiwan citizens. They grouped all factors into three groups of "factors related to mobile", "factors related to mobile hardware" and "psychological factors of users" [51]. However, these studies had two major shortcomings. The first of which was that the cause and effect relationships among factors affecting technology adoption were not considered and the second one was that the intensity of these relationships were not also taken into consideration.

To tackle above mentioned limitations, this study has classified critical factors of technology adoption into two groups of general group and specific group. The general group includes commonly-held factors that are usually discussed in literature of technology adoption. Specific factors group comprises some typical factors influencing a specific type of technology that the Long Term Evolution (LTE) is the case of this study. then, all general and specific factors are listed in an integrated framework and a fuzzy DEMATEL methodology has been used to identify critical influencers of mobile networks technology adoption. The fuzzy nature of the model is because of the fact that the effect value of each factor is based on lingual judgements and such judgements include a level of ambiguity and imprecision. Therefore, the fuzzy logic has been used to assess the such judgements in a better way. The major contributions of this paper can be (1) developing a framework for assessing the critical factors of the fourth generation of mobile networks (4G) technology (2) visualization of cause and effect relationships among these factors and (3) extraction and identification of critical success factors of 4G adoption in Iranian society. The rest of this paper is organized as following the second section deals with literature review. In this section, general and specific theories of technology adoption are discussed. The research methodology is explicated in the third section. Sampling techniques, data gathering methodology and DEMATEL and fuzzy logic are viewed in this section. The fourth section discusses analysis of factors of 4G technology adoption, extraction of causal and effect group. The fifth and last section ends with a conclusion and some directions for future researches.

II. LITERATURE REVIEW

*General theories of technology adoption*

Diffusion of innovation theory (DIT) proposed by Rogers in 1960s is believed to have had a very profound effect over studies of innovation adoption and diffusion. This theory discusses four factors affecting diffusion of an innovation including (1) the innovation, (2) communication channels, (3) time and (4) social setting [39] DIT is regarded as one of the most outstanding frameworks of innovation diffusion study. In the late 1970s, Fishbein and Ajzen developed theory of reasoned action (TRA) which rooted in social psychology. This theory has 3 general constructs of 1) behavioral intention, 2) attitude and 3) subjective norm. according to TRA, the behavioral intention of a person depends on his or her attitude and subjective norms. It can be interpreted that the behavioral intention is the resultant of attitude and subjective norms. Moreover, if one has a sufficiently powerful intention for doing a specific behavior, the intention is very likely to lead to that behavior [40]. Theory of planned behavior (TPB) was proposed by Fishbein and Ajzen in 1975. TPB was an extension of TRA through adding the construct of perceived behavioral control to it. This construct deals with perception of people about ease or difficulty if doing their desired behavior. One of the basic critiques to TRA is that it is built upon somewhat static constructs of motivation and can't be used to forecast behavioral outcomes. TRA is actually rooted in self-efficacy theory proposed by Bandura that it itself is derived from social cognitive theory. So, self-efficacy is the most important determinant of behavioral change because it leads to the development of managerial behavior [41].

Technology acceptance model (TAM) is widely used due to its simple structure. TAM was first proposed by Davis in 1989 for researching in social psychology. This model is still in use by a number of researchers. TRA and TPB are two basic theories of social psychology underlying TRA. It claims adoption of a technology depends on its perceived ease of use and perceived usefulness. Perceived ease of use of a technology is a degree to which a person finds the use of that technology easy and facile. Perceived usefulness is a degree to which a person thinks the application of a technology can increase his or her performance. According to TAM, the perceived usefulness of a technology depends on its perceived ease of use because the easier a person finds a technology the more useful her or she considers it [42].

Thompson and associates (1991) proposed personal computer utilization model (PCUM) through studying the conditions driving the adoption of PCs by 212 organizational knowledge workers in nine divisions of a multi-national corporations. Though this model is often used in IT domain, it is taken as a general theory of technology adoption due to its very high application. It shows the tendency of a person towards adoption of a technology depends on several factors such as his or her affection about that technology, contextual factors, long-term expected payoffs of its use, complexity and job relevance [44]. Davis and colleagues (1992) used a motivation model (MM) to study IT technology adoption. According to MM, the behavior is driven by extrinsic and intrinsic motives. The extrinsic motive is defined as a perception that a person wants to perform an action. Some examples of this perception entails perceived usefulness such as reward and promotion, perceived ease of use and social norm. in addition, if an action results in the enjoyment and satisfaction of a person, it can be considered as an intrinsic motive. Enjoying computer games can be an example of intrinsic motives [45].

Venkatesha and Davis (2000) extended TAM (TAM2) through extracting the determinants of perceived usefulness. These factors comprised social influence processes like subjective norms and images and cognitive instrumental processes such as job relevance, output quality and result demonstrability [43]. Venkatesh et al. (2003) proposed the unified theory of acceptance and use of technology (UTAUT). It was actually an integrated framework of diffusion of innovation theory (DIT), theory of reasoned action (TRA), theory of planned behavior (TPB), technology acceptance model (TAM), model of personal computer utilization (MPCU), motivation model (MM), extended technology acceptance model (TAM2). This model includes four basic constructs of performance expectancy, effort expectancy, social influence and facilitating conditions. Besides, there are four moderating variables of age, gender, experience and voluntary use each of which affects some of model's basic constructs. These four dimensions play a critical role in adoption of a technology. In contrast to previous models that could explain the 30% of adopter's behavior, UTAUT could explain it with 70% accuracy.

*Specific theories of technology adoption*

Boyd and Mason (1999) in a study indicated that factors such as company's reputation and credibility affect the adoption of its innovation [23]. Keaveney and Parthasasrathy (2001) in a study on the tendency of people toward online services investigated the effects of mental, behavioral and demographic factors and showed that factors such as advertising, media and press are of great importance in the adoption of such services [24]. Sarker and Wells (2003) in a study on the adoption and use of mobile devices, investigated the role of factors such as the impact of peer groups, friends and relatives' recommendations, the number of users and (perceived) usefulness on the adoption of this technology [25]. Teo and Pok (2003) in a study on mobile handset adoption explored the role of three factors of the peer groups, friends and relatives' recommendations and the number of users [26]. Pagani (2004) in a study in relation to the determinant factors in adoption of third generation mobile networks showed that factors such as handset price, data transmission cost in the network, network coverage, signal quality, transmission speed, perceived usefulness, perceived ease of use and device adaptability with network are among effective factors in third generation network adoption. Yang (2005) in a study that identified the effective role of adoption of mobile commerce in Singapore, considered the role of factors such as (perceived) usefulness and ease of use [27]. Social environment includes social factors, facilitating conditions and social impact. Friends and relatives' recommendations include the impact of peer groups. Professional relation involves job fit. Signal quality (output) includes output quality. Attitude includes affection toward technology, behavioral intentionality for use and subjective norms.

*Table 2: Summary of literature on effective factors on technology adoption*

| Factors affecting adoption of mobile services technology | Rogers (1960) | Fishbein and Ajzen (1977) | Ajzen (1985) | Davis (1989) | Tompson et al. | Davis et al. | Boyd and Mason (1999) | Ventaketsh and Davis | Venkatesh et al. (2003) | Keaveney and Parthasasthy (2001) | Sarker and Wells (2003) | Teo and Pok (2003) | Pagani (2004) | Yanh (200) | Lu et al. (2005) | Lu et al. (2005) | Scheepers et al. (2006) | Blechar et al. (2006) | Ranganathan et al. (2006) | Schierz et al. (2010) | Hill and troshani (2010) | Lin (2011) | Chong et al (2011) | Shuchiung et al. (2015) |
|---|---|---|---|---|---|---|---|---|---|---|---|---|---|---|---|---|---|---|---|---|---|---|---|---|
| Phone price (handset) | | | | | | | | | | | | | * | | | | | | | | | | | |
| Transmission cost (call cost) | | | | | | | | | | | | | * | | | | | | | | | | | |
| Network coverage | | | | | | | | | | | | | * | | | | | | | | | | | |
| Communication channels (press and media | * | | | | | | | | | | | | | | | | | | | | | | | |
| Friends and relatives' recommendations | | | | | | | | | | * | * | | | | * | | * | | | | | | * | |
| number of users | | | | | | | | | | * | * | | | | * | | | * | | | | | * | |
| Security and privacy | | | | | | | | | | | | | | | * | | | | | | | | | |
| Signal quality | | | | | | | | * | | | | | * | | | | | | | | | | | |
| Transmission speed | | | | | | | | | | | | | * | | | | | | | | | | * | * |
| Usefulness (perceived usefulness) | | | * | * | * | * | | | * | | * | * | * | * | * | | | * | | * | * | * | * | * |
| Perceived ease of use | | | | * | * | * | | | * | | | * | | | | | | | | * | | * | * | |
| Device (phone)adaptability | | | | | | | | | | | | | * | | * | | | | | | * | | | |
| Company's reputation (validity) | | | | * | | * | | | | | | | | | * | | | | * | | | | | |
| trust | | | | | | | | | | | | | | | | | | | | | | | | * |
| uncertainty | | | | | | | | | | | | | | | | | | | | | | | | * |
| innovation itself (technical superiority ) | * | | | | | | | | | | | | | | | | | | | | | | | |
| advertisement | | | | | | | | | | | * | | | | | | * | | | | | | | |
| time | * | | | | | | | | | | | | | | | | | | | | | | | |
| Social environment | * | | | * | * | | | | * | | | | | | | | | | | | | | | |
| attitude | | * | * | * | * | * | | * | | | | | | | | | | | | | | | | |
| Perceived behavioral control | | | * | | | | | | | | | | | | | | | | | | | | | |
| Professional relationship | | | | | | | | | * | | | | | | | | | | | | | | | |
| Result demonstrability | | | | | | | | * | * | | | | | | | | | | | | | | | |
| Long term outcomes | | | | | | | | * | | | | | | | | | | | | | | | | |
| Performance expectance | | | | | | | | | * | | | | | | | | | | | | | | | |
| Effort expectance | | | | | | | | | * | | | | | | | | | | | | | | | |
| gender | | | | | | | | | * | | | | | | | | | | | | | | | |
| age | | | | | | | | | * | | | | | | | | | | | | | | | |
| experience | | | | | | | | | * | | | | | | | | | | | | | | | |

Yang (2005) in a study in relation to identifying effective factors on mobile commerce adoption in Singapore, indicated that factors such as (perceived) usefulness and ease of use have direct impact on this subject [28]. In a study conducted in relation to adoption of wireless internet services via mobile, it was shown that factors such as peer groups, friends and relatives' recommendations, the number of users and perceived usefulness affect the adoption of wireless internet services via mobile [29]. Lu, et al (2005) investigated the role of factors such as advertising (media and press), security and privacy, handset adaptability, reputation (face), company's (service provider) validity in adoption of mobile technology [30]. In a study on how the environment affects users' satisfaction of mobile computing services, Scheepers et.al (2006) studied the role of factors such as the impact of peer groups and friends and relatives' recommendations [31]. The number of users and perceived usefulness are among the factors studied by Blechar et.al in 2006 [32]. Company's reputation and prestige were among the factors considered by Ranganathan et.al in 2006 [33]. Perceived usefulness and ease of use were among the factors investigated by Schierz et.al in 2010 [34].

Hill and Troshani (2010) in their study on effective factors on mobile services adoption, analyzed the role of perceived usefulness. Lin (2011) in an empirical study on factors affecting mobile banking adoption, investigated the impact of different factors such as perceived usefulness, ease of use, handset adaptability on mobile services adoption [36]. Chong et.al (2011) studied the effect of factors such as peer groups, friends and relatives' recommendations, the number of users, transmission speed, usefulness and ease of use on adoption of mobile learning in Malaysia [37]. Shu-Chiung et.al (2014) studied the adoption of fourth-generation technology from the point of view of the two approaches of technology appropriateness and the cost of economic relations. Technology appropriateness approach is exactly based on TAM and the theory of economic relations cost is composed of two structures of trust and uncertainty. This study indicated that, apart from the TAM-constituent structures, the cost of economic interactions is also a very important factor in adoption of the fourth-generation network. So that with the increase of trust, the user's desire to use the fourth-generation network is increased too (positive correlation), however the increase in uncertainty has a great effect on reducing the user's desire to use the fourth generation network (negative correlation) [38]. The summary of literature review is presented in

Table 2 .

*Critical success factors*

Each system consists of several factors, and management and system improvement depends on the correct recognition of these factors. Since the number of factors, affecting a system is large, recognition of critical effective factors or critical success factors (CSF) is of great importance. The concept of CSFs was first proposed by John F. Rockart (1979) from the MIT School of Management in Slovan. Rockart describes CSFs as a key area of the activity of an organization, the desirable status of which is necessary for the fulfillment of organization's goals. CSFs include conditions, attributes, or variables that have to go well for the success of any organization and institution [13]. Thierauf believes that if CSFs status is not appropriate, organizations will fail to meet their goal. And when CSFs are in an appropriate status, the organization will develop and achieve its goal [14]. Among the highly useful methods for identifying CSFs, case studies, expert and decision makers' interviews [15] and the Delphi method can be referred to [3]. However, since these methods are simply influenced by the mentality of individuals, it is better to use multi-criteria decision-making methods to identify CSFs. In this study Decision Making Trial and Evaluation laboratory method (DEMATEL) is used. This method which was first proposed by Gabus and Fontela 1979 [16-17] allows analysts to gather group knowledge from specialists and transform them into a structured model, and then show cause-effect relationships between factors through the cause-diagram. DEMATEL outputs specify information about the influence that each factor has on the whole system. By analyzing the structural model, one can find out what factors are more fundamental to the system and which factors are less fundamental to the system. As a result, the influential factors (which have the highest impact on the system) are considered critical factors of success [15].

III. MATEERIALS AND METHDOLOGY

*Sampling*

The statistical population of this study is comprised of telecommunication experts, through interviewing whom, effective factors on technology adoption were identified (components of mental model). This sampling was performed by non-random targeted sampling method [54]. A targeted sampling technique is a type of sampling in which specific situations, individuals, or events are mainly chosen because of the important information they can provide, so that such information cannot be obtained from other options [55]. Telecommunication technology experts are selected as individuals who have useful information about the factors influencing adoption of mobile technologies. These experts were selected from the major telecommunication organizations of I.R. Iran, including Telecommunication Infrastructure Company of I.R. Iran, Communication Regulatory Authority of I.R. Iran, Information Technology Organization of I.R. Iran, I.R. Iran Telecommunication Research Center, Mobile Telecommunications Company of Iran (MCI), MTN Irancell (Second Operator), Rightel (Third Operator). It should be noted that according to the nature of the subject under study, the interview was conducted only with a group of people who still had not used the fourth-generation technology. The statistical characteristics of the first sample respondents are as shown in Table 3.

*Table 3: Demographic information of respondents (sample = X)*

| Demographic Information | | Frequency | Percentage |
|---|---|---|---|
| Gender | Male | 35 | 0.7 |
| | Female | 15 | 0.3 |
| Age range | 20-30 | 9 | 0.18 |
| | 31-40 | 19 | 0.38 |
| | 41-50 | 17 | 0.34 |

|  | Over 50 | 5 | 0.1 |
|---|---|---|---|
| Education | Undergraduate | 9 | 0.18 |
|  | Bachelor | 16 | 0.32 |
|  | Master | 21 | 0.42 |
|  | PhD | 4 | 0.08 |
| Position | Expert | 13 | 0.25 |
|  | Senior Expert | 22 | 0.44 |
|  | Assistant director of general manager | 9 | 0.18 |
|  | General Manager | 6 | 0.12 |
| Organization | Telecommunication Infrastructure Company | 5 | 0.1 |
|  | Communication Regulatory Authority | 9 | 0.18 |
|  | Information Technology Organization | 5 | 0.1 |
|  | Iran Telecommunication Research Center | 6 | 0.12 |
|  | Mobile Telecommunications Company of I Iran (MCI) | 9 | 0.18 |
|  | MTN Irancell | 9 | 0.18 |
|  | Rightel | 7 | 0.14 |

*Data gathering methodology*

Since the extraction of the importance of each of the factors required a direct discussion with experts, in this research, interviewing method was used for data collection. 29 factors were investigated in this study. Accordingly, 841 questionnaires were prepared in 6 separate questionnaires. Five 145- item questionnaires and a 116-item questionnaire were discussed with six experts in six sessions.

*Modeling methodology*

DEMATEL is a very effective tool for gathering opinions from system experts, analyzing the interactions between components of a system and displaying them in a cause-effect structure. This method proposed by Gabus and Fontela in the early 1970's [17-16], allows analysts to gather knowledge of a group of experts and transform them into a structured model, and then show the cause-effect relationships between the factors through cause diagram. The evaluation of complex systems by experts always has some ambiguity and uncertainty, so that the respondents are interested in applying linguistic judgments rather than using definite numerical criteria [06, 58, 57]. To overcome this problem, fuzzy logic has been used [59]. In this paper, by combining fuzzy logic and DEMATEL's methodology, a systematic framework for identifying the critical factors affecting the adoption of fourth-generation technology in. Iranian society has been presented, which, as authors know, is the first study to be done in this regard.

*DEMATEL*

DEMATEL is a method for analyzing complex systems. This method, which is based on the theory of graph, enables researchers to identify the interdependencies between the components of the system, analyze their severity and display them in a cause-effect relationship. In this method, the constituent elements of the system are divided into two groups: cause and effect. Thus, researchers can more clearly distinguish the structural relations between the constituent elements of the system [61,62,63]. One of the features of DEMATEL is the ability of this method to visualize the interactions between the constituent elements of the system in a cause and effect format. Today, this method is being used in a wide variety of domains.

In a study conducted by Wu and Lee in 2007, DEMATEL is used to identify and classify competitive advantages into different groups, each of which could be exclusively developed and improved. Using the concept of multiple intelligences, they divide all the core competencies into eight different IQs called cognitive IQ, emotional IQ, political IQ, cultural / social IQ, organizational IQ, network IQ, innovative IQ and intuitive IQ. They showed that cognitive IQ, emotional IQ, innovative IQ and intuitive IQ are four critical factors that are very necessary for the development of global capabilities of managers. In their study, in order to resolve the ambiguity in expert judgments, they completed DEMATEL through fuzzy logic [65].

In 2008, by studying the constituent elements of a system, Wu could assess the effective strategies of implementing knowledge management by mixing DEMATEL and the network analysis process (ANP). In this study, three general clusters, named targets cluster, criteria cluster and strategies cluster were investigated. The target cluster included information activation, performance improvement and innovation promotion. The criteria cluster included things such as incentives, high-level organization management support, time, cost, culture, and communications. The cluster of strategy also included coding strategies, individualization strategies, and hybrid strategies. The results of this study identified information activation as the most important goal and individualization strategy as the most effective strategy. [64]. In a research, Zhu et al. used the Fuzzy DEMATEL method to identify the critical factors affecting emergency management. In this study, 20 factors influencing emergency management were studied. Using the DEMATEL method, these twenty factors were divided into two groups of cause and effect groups, and according to the net effect of each of the factors, five critical factors were extracted from them. [61]

Sheih et al. in a study considered the critical factors in the quality of hospital services. In this study, first by using the SERVQUAL model, seven major factors in the quality of hospital services were extracted. These factors included "modern medical equipment," "the communication ability of the service personnel," "trusted and expert medical staff," "service personnel with problem solving skills", "detailed description of the patient's condition by the physician", "medical staff with professional skills" and "pharmacists' recommendations for drug use". Finally, using the DEMATEL method, the first four factors were identified as the critical factors affecting hospital services [70].

Mirmousa and Dehghan Dehnavi in a study using the Fuzzy Delphi and Fuzzy DEMATEL methods, identified and extracted the critical factors influencing the selection of suppliers. First, by studying a large number of studies, 44 effective criteria were extracted in the selection of suppliers. In the next step, 14 factors were extracted from them using Fuzzy DEMATEL. In the final step, the Fuzzy DEMATEL method was used to identify the most important factor. The results of this study showed that the financial stability criterion is the most important factor in choosing the supplier [71].

According to the definition of Gabus and Fontela (16-17), the design of the DEMATEL consists of five steps as follows:

Step 1: Creating the initial direct relationship matrix $[A = [aij]]$: For this purpose, set up a team of experts and ask their judgments about the direct impact of each pair of factors. These judgments can be obtained through a questionnaire or in-depth interview. They can be in the form of exact numbers or verbal judges(terms). Since the evaluation involves some ambiguity and uncertainty, so as to overcome such problem, using verbal judgments is more appropriate. After obtaining verbal judgments, they must be converted into definite values.

$A = [a_{ij}]$ is a non-negative matrix N×N, where $a_{ij}$ indicates the effect of the factor $i$ on the factor $j$ and when $i = j$, the elements of the original diameter are $0 = aij$.

Step 2: non-scalability of the Initial Direct Relationship matrix by the relationship (1). All elements in the matrix D are valid at intervals $0 \leq d_{ij} \leq 1$, and the elements of the original diameter are zero.

$$D = \frac{1}{\max_{1 \leq i \leq n} \sum_{j=1}^{n} a_{ij}} A \qquad (1)$$

Step 3: Obtaining T general relation matrix by equation (2), so that I is the same matrix. The element $T_{ij}$ represents an indirect relationship in which i effects j. Thus, the matrix T represents the general relationship between each pair of system factors.

$$T = D(I - D)^{-1} \qquad (2)$$

Step 4: Calculating the influential and influenced degree of each factor through (3) and (4). The influential and influenced degree of the factors and identification of cause group and effect factors group are characterized by:

$$r_i = \sum_{1 \leq j \leq n} t_{ij} \qquad (3)$$

$$c_j = \sum_{1 \leq i \leq n} t_{ij} \qquad (4)$$

Sum of the line i, shown as ri, denotes all the direct and indirect effects that factor i has on all other factors, and hence i can be called influential degree. Similarly, the line j represented as cj can be called the influenced degree, since cj is the sum of all the direct and indirect effects that factor j receives from other factors.

Therefore, it is natural that when i = j, ri + ci represents all the effects received by the factor i. In other words, ri + ci represents the effect of factor i on the whole system and the effect that other system factors have on the factor i. Thus, the ri + ci index can indicate the importance degree that the factor i has in the system as a whole. In contrast, the difference between these two ri-ci represents the net effect that the factor i has on the whole system. Specifically, if the value of ri-ci is positive, the factor i is a net cause and belongs to the group of causes, but if the value of ri-ci is negative, the factor i is a net effect and belongs to the effect group.

*Fuzzy logic*

Organizational managers use group decision making as a tool for making satisfactory decisions. Group decision making is a result of the opinions of all the participating experts. When issues of the organization become wider and more complex, the evaluations provided by the experts are expressed in terms of verbal statements rather than exact and definite values. These evaluations and verbal statements are subject to ambiguity and uncertainty, which makes the decision-making process more difficult. A fuzzy theory developed based on fuzzy sets provides a framework for managing and resolving ambiguities and uncertainties in decision making. Fuzzy theory has been used in many areas such as artificial intelligence, automatic control, image recognition, medical diagnosis, psychology, management science, weather forecasting and environmental assessment, and it has shown very good results [66]. In fuzzy logic, each number between zero and one is considered to be almost correct (true), while in definite sets, (two-value sets) everything is either correct (one) or incorrect (zero). Fuzzy logic enables researchers to evaluate ambiguous and inaccurate judgments. With the help of fuzzy logic, vague verbal judgments can be converted to fuzzy numbers. Some of the most common fuzzy numbers include triangular fuzzy numbers, trapezoidal fuzzy numbers, and Gaussian fuzzy numbers. Table 4 is a correspondence of verbal judgments and fuzzy numbers.

*Table 4: Correspondence of verbal judgments and related triangular fuzzy numbers*

| term | Fuzzy number |
| --- | --- |
| No effect | (0 0 0.25) |
| little effect | (0 0.25 0.5) |
| medium effect | (0.25 0.5 0.75) |
| high effect | (0.5 0.75 1) |
| Very high effect | (0.75 1 1) |

The C-triangular fuzzy number can be defined as threefold $(l, m, r)$, so that we have $l \ll m \ll r$. For the two fuzzy numbers $c_1 = \{l_1, m_1, r_1\}$ and $c_2 = \{l_2, m_2, r_2\}$ the following algebraic operations can be defined:

$$C_1 + C_2 = (l_1 + l_2, m_1 + m_2, r_1 + r_2)$$

$$C_1 - C_2 = (l_1 - l_2, m_1 - m_2, r_1 - r_2)$$

$$C_1 \otimes C_2 = (l_1 l_2, m_1 m_2, r_1 r_2)$$

$$C_1 \div C_2 = (l_1 \div l_2, m_1 \div m_2, r_1 \div r_2)$$

$$\lambda C_1 = (\lambda l_1, \lambda m_1, \lambda r_1)$$

Because fuzzy numbers cannot be used directly for matrix operations, it is necessary to defuzzify them. Defuzzification allows researchers to transform fuzzy numbers into definite numbers and perform matrix operations on them. There are many defuzzification methods, some of which include the centroid method, the mean value method using left and right separations [11], the method of converting fuzzy data to crisp score (CFCS) [19], etc. For a detailed study of fuzzy methods, the reader can refer to resources [22,12,20]. The main objection to the centroid method as one of the most widely used defuzzification methods is that this method cannot separate two disjoint (with different forms) fuzzy numbers that can be converted into similar definite number. Therefore, the CFCS method is used here (more reasons are needed). The CFCS process is as follows [19]:

In the first step, based on the fuzzy numbers obtained from the expert's opinion, the left and right values are extracted from the fuzzy mean and fuzzy max. Assume that $\tilde{A}_{ij}^k = (l_{ij}, m_{ij}, r_{ij})$ and $1 \ll k \ll K$ equals to the K-expert's fuzzy assessment from the effect of factor i on factor j. In the second step, fuzzy numbers are standardized:

$$xl_{ij}^k = (l_{ij}^k - \min_{1 \le k \le K} l_{ij}^k)/\Delta_{min}^{max} \qquad (5)$$

$$xm_{ij}^k = (m_{ij}^k - \min_{1 \le k \le K} l_{ij}^k)/\Delta_{min}^{max} \qquad (6)$$

$$xr_{ij}^k = (r_{ij}^k - \min_{1 \le k \le K} l_{ij}^k)/\Delta_{min}^{max} \qquad (7)$$

Where we have

$$\Delta_{min}^{max} = max\ r_{ij}^k - min\ l_{ij}^k \quad (8)$$

In step three, the normalized left and right values are calculated as follows

$$xls_{ij}^k = xm_{ij}^k/(1 + xm_{ij}^k - xl_{ij}^k) \quad (9)$$

$$xrs_{ij}^k = xr_{ij}^k/(1 + xr_{ij}^k - xm_{ij}^k) \quad (10)$$

In the fourth step, all normalized values are calculated:

$$x_{ij}^k = [xls_{ij}^k(1 - xls_{ij}^k) + xrs_{ij}^k xrs_{ij}^k]/(1 + xrs_{ij}^k - xls_{ij}^k) \quad (11)$$

In the fifth step the definite score of K expert is calculated as follows:

$$BNP_{ij}^k = min\ l_{ij}^k + x_{ij}^k \Delta_{min}^{max} \quad (12)$$

In the sixth step, the integrated score ( best non fuzzy performance)of all the experts is calculated as follows:

$$a_{ij} = \frac{1}{k} \sum_{k}^{1 \leq k \leq K} BNP_{ij}^k \quad (13)$$

*Development of fuzzy DEMATEL*

The process of designing fuzzy DEMATEL is as follows:
1: Calculation of the fuzzy matrix of each expert
$\tilde{A}_k = [\tilde{a}_{ij}]_k$: The views of each expert $E_k, k = 1, \ldots \ldots K$ about the effect of factor i on the factor j are extracted through a fuzzy questionnaire. The terms (verbal judgments) and fuzzy triangular numbers corresponding to each is shown in Table (4).
2: Creating a combined fuzzy matrix ($\tilde{A} = [\tilde{a}_{ij}]$): The combined fuzzy matrix, which is the result of the aggregation and average of all matrices of experts, is calculated by the following equation:

$$\tilde{A} = \frac{1}{K} \sum_{k=1}^{K} \tilde{A}_k \quad (14)$$

3: Converting ($\tilde{A}$) to the definite matrix ($A = [a_{ij}]$) through CFCS defuzzification methods.
4: Normalization of the definite matrix ($A = [a_{ij}$) and creating matrix ($D = [d_{ij}]$), matrix ($D = [d_{ij}]$) are calculated through relation (1).
5: The total relation matrix ($T = [t_{ij}]$): matrix of total relations($T$) is calculated based on relation (2):
6: Calculating influential and influenced degree of each of the factors: The influential and influenced degree of each of the factors and the identification of the cause and effect factor group are specified through relations (3) and (4):
Sum of the line i, shown as ri, denotes all the direct and indirect effects that factor i has on all other factors, and hence, i can be called degree of influential impact. Similarly, the line j represented as cj can be called degree of influenced impact because cj is the sum of all the direct and indirect effects that factor j receives from other factors. Therefore, it is natural that when i = j, ri + ci represents all the effects received by the factor i. In other words, ri + ci also indicates the effect that factor i has placed on the system and the effect that other system factors have on factor i. Therefore, the ri + ci index can indicate the importance of the factor i in the whole system. In contrast, the difference between these two ri-cis represents the net effect that the factor i has on the whole system. In particular, if the value of ri-ci is positive, factor i is a net cause (group of causes). But if the value of ri-ci is negative, factor i is a net effect (effect group).

**7.** drawing a cause-effect diagram

The cause and effect diagram is obtained by mapping the dataset (ri-ci, ri + ci). The complex relationships between the factors are created through the process of creating the diagram. It is clear that the cause and effect groups can be derived from this diagram.

IV. ANALYSIS OF CSFs OF TECHNOLOGY ADOPTION

*Application of proposed model*

After identifying and extracting the factors affecting the adoption of the fourth generation technology through studying the literature (

Table 2), it is now necessary to extract the critical factorsaffecting the adoption of the fourth generation technology through the fuzzy DEMATEL method. Since 29 factors were extracted from the literature, at first, a questionnaire was developed in the form of 841 questions, which were discussed during 6 sessions with experts.

To answer questions, the scale presented in Table 4was used. Using the CFCS method, all these verbal evaluations are aggregated into definite values, which represent the amount of direct effect that each factor has over other system factors. In this way, the initial direct relation matrix is presented in Table 5. Then, using equation (2), the total relation matrix is presented in

Table 6. Using the general relation matrix, the score of all factors is calculated by the equations (3) and (4) and are shown in the

Table 7 Based on the results obtained in this table, the cause and effect relationship between all the factors are shown in Figure 5.

*Table 5 :the initial direct relations*

| | X1 | X2 | X3 | X4 | X5 | X6 | X7 | X8 | X9 | X10 | X11 | X12 | X13 | X14 | X15 | X16 | X17 | X18 | X19 | X20 | X21 | X22 | X23 | X24 | X25 | X26 | X27 | X28 | X29 |
|---|---|---|---|---|---|---|---|---|---|---|---|---|---|---|---|---|---|---|---|---|---|---|---|---|---|---|---|---|---|
| X1 | 9.7 | 10.16 | 9.73 | 10.1 | 11.79 | 12.16 | 10.8 | 10.4 | 10.4 | 11.8 | 11.19 | 9.796 | 12.01 | 11 | 11 | 10.5 | 10.9 | 9.46 | 11.11 | 12.46 | 9.6 | 10.9 | 10.7 | 10.8 | 11.8 | 11.72 | 9.5 | 9.66 | 11.1 |
| X2 | 8.4 | 8.867 | 8.43 | 8.79 | 10.5 | 10.87 | 9.5 | 9.01 | 9.07 | 10.5 | 9.9 | 8.5 | 10.72 | 9.7 | 9.67 | 9.2 | 9.63 | 8.17 | 9.822 | 11.17 | 8.3 | 9.57 | 9.4 | 9.48 | 10.5 | 10.43 | 8.2 | 8.37 | 9.83 |
| X3 | 7.8 | 8.3 | 7.87 | 8.23 | 10 | 10.4 | 9.07 | 8.51 | 8.5 | 9.97 | 9.4 | 7.933 | 10.28 | 8.4 | 8.42 | 8.767 | 9.15 | 7.6 | 9.356 | 10.73 | 7.7 | 9.07 | 8.83 | 8.91 | 10.1 | 10 | 7.63 | 7.8 | 9.27 |
| X4 | 8 | 8.467 | 8.03 | 8.4 | 10.1 | 10.47 | 9.1 | 8.61 | 8.67 | 10.1 | 9.5 | 8.1 | 10.32 | 9.2 | 9.23 | 8.8 | 9.23 | 7.77 | 9.422 | 10.77 | 7.9 | 9.17 | 9 | 9.08 | 10.1 | 10.03 | 7.8 | 7.97 | 9.43 |
| X5 | 6.8 | 7.267 | 6.83 | 7.21 | 8.9 | 9.267 | 7.9 | 7.41 | 7.47 | 8.87 | 8.3 | 6.9 | 9.117 | 8 | 8.02 | 7.6 | 8.03 | 6.57 | 8.222 | 9.567 | 6.7 | 7.97 | 7.8 | 7.88 | 8.9 | 8.833 | 6.6 | 6.77 | 8.23 |
| X6 | 10 | 10.47 | 10 | 10.4 | 12.1 | 12.47 | 11.1 | 10.6 | 10.7 | 12.1 | 11.5 | 10.1 | 12.32 | 11 | 11.3 | 10.8 | 11.2 | 9.77 | 11.42 | 12.77 | 9.9 | 11.2 | 11 | 11.1 | 12.1 | 12.03 | 9.8 | 9.97 | 11.4 |
| X7 | 11 | 11.47 | 11 | 11.4 | 13.1 | 13.47 | 12.1 | 11.6 | 11.7 | 13.1 | 12.5 | 11.1 | 13.32 | 12 | 12.3 | 11.8 | 12.2 | 10.8 | 12.42 | 13.77 | 11 | 12.2 | 12 | 12.1 | 13.1 | 13.03 | 10.8 | 11 | 12.4 |
| X8 | 14 | 14.07 | 13.6 | 14 | 15.7 | 16.07 | 14.7 | 14.2 | 14.3 | 15.7 | 15.1 | 13.7 | 15.92 | 15 | 14.9 | 14.4 | 14.8 | 13.4 | 15.02 | 16.37 | 14 | 14.8 | 14.6 | 14.7 | 15.7 | 15.63 | 13.4 | 13.6 | 15 |
| X9 | 14 | 14.07 | 13.6 | 14 | 15.7 | 16.07 | 14.7 | 14.2 | 14.3 | 15.7 | 15.1 | 13.7 | 15.92 | 15 | 15 | 14.4 | 14.8 | 13.4 | 15.02 | 16.37 | 14 | 14.8 | 14.6 | 14.7 | 15.7 | 15.63 | 13.4 | 13.6 | 15 |
| X10 | 9.3 | 9.794 | 9.36 | 10 | 11.41 | 11.77 | 10.4 | 9.94 | 9.99 | 11.4 | 10.81 | 9.428 | 11.63 | 11 | 10.6 | 10.13 | 10.6 | 9.09 | 10.74 | 12.07 | 9.2 | 10.5 | 10.3 | 10.4 | 11.4 | 11.36 | 9.13 | 9.29 | 10.8 |
| X11 | 10 | 10.67 | 10.2 | 10.6 | 12.3 | 12.67 | 11.3 | 10.8 | 10.9 | 12.3 | 11.7 | 10.3 | 12.52 | 12 | 11.5 | 11 | 11.4 | 9.97 | 11.62 | 12.97 | 10 | 11.4 | 11.2 | 11.3 | 12.3 | 12.23 | 10 | 10.2 | 11.6 |
| X12 | 9 | 9.467 | 9.03 | 9.4 | 11.1 | 11.47 | 10.1 | 9.61 | 9.67 | 11.1 | 10.5 | 9.1 | 11.32 | 10 | 10.4 | 9.8 | 10.2 | 8.77 | 10.42 | 11.77 | 8.9 | 10.2 | 10 | 10.1 | 11.1 | 11.03 | 8.8 | 8.97 | 10.4 |
| X13 | 9 | 9.467 | 9.03 | 9.42 | 11.1 | 11.47 | 10 | 9.61 | 9.67 | 11.1 | 10 | 9.1 | 11.32 | 10 | 10.3 | 9.8 | 10.2 | 8.77 | 10.42 | 11.77 | 8.9 | 10.2 | 10 | 10.1 | 11.1 | 11.03 | 8.8 | 8.97 | 10.4 |
| X14 | 9.4 | 9.867 | 9.43 | 9.81 | 11.5 | 11.87 | 10.5 | 10 | 10.1 | 11.5 | 10.9 | 9.5 | 11.72 | 11 | 10.7 | 10.2 | 10.6 | 9.17 | 10.82 | 12.17 | 9.3 | 10.6 | 10.4 | 10.5 | 11.5 | 11.43 | 9.2 | 9.37 | 10.8 |
| X15 | 10 | 10.47 | 10 | 10.4 | 12.1 | 12.47 | 11.1 | 10.6 | 10.7 | 12.1 | 11.5 | 10.1 | 12.32 | 11 | 11.4 | 10.8 | 11.2 | 9.77 | 11.42 | 12.77 | 9.9 | 11.2 | 11 | 11.1 | 12.1 | 12.03 | 9.8 | 9.97 | 11.4 |
| X16 | 18 | 18.5 | 18.1 | 18.4 | 20.2 | 20.6 | 19.3 | 18.7 | 18.7 | 20.2 | 19.6 | 18.13 | 20.48 | 18 | 18 | 18.97 | 19.4 | 17.8 | 19.56 | 20.93 | 18 | 19.3 | 19 | 19.1 | 20.3 | 20.2 | 17.8 | 18 | 19.5 |
| X17 | 11 | 11.3 | 10.9 | 11.2 | 13 | 13.4 | 12.1 | 11.5 | 11.5 | 13 | 12.4 | 10.93 | 13.28 | 11 | 11.2 | 11.77 | 12.2 | 10.6 | 12.36 | 13.73 | 11 | 12.1 | 11.8 | 11.9 | 13.1 | 13 | 10.6 | 10.8 | 12.3 |
| X18 | 9.4 | 9.9 | 9.47 | 9.84 | 11.6 | 12 | 10.7 | 10.1 | 10.1 | 11.6 | 11 | 9.533 | 11.95 | 9.9 | 9.94 | 10.37 | 10.8 | 9.2 | 10.96 | 12.33 | 9.3 | 10.7 | 10.4 | 10.4 | 11.7 | 11.6 | 9.23 | 9.4 | 10.9 |
| X19 | 9.6 | 10.07 | 9.63 | 10 | 11.7 | 12.07 | 10.7 | 10.2 | 10.3 | 11.7 | 11.1 | 9.7 | 11.92 | 11 | 10.9 | 10.4 | 10.8 | 9.37 | 11.02 | 12.37 | 9.5 | 10.8 | 10.6 | 10.7 | 11.7 | 11.63 | 9.4 | 9.57 | 11 |
| X20 | 9.2 | 9.667 | 9.23 | 9.62 | 11.3 | 11.67 | 10.3 | 9.81 | 9.87 | 11.3 | 10.7 | 9.3 | 11.53 | 10 | 10.5 | 10 | 10.4 | 8.97 | 10.64 | 11.97 | 9.1 | 10.4 | 10.2 | 10.3 | 11.3 | 11.23 | 9 | 9.17 | 10.6 |
| X21 | 1.4 | 1.867 | 1.43 | 1.82 | 3.5 | 3.867 | 2.5 | 2.01 | 2.07 | 3.47 | 2.9 | 1.5 | 3.717 | 2.6 | 2.62 | 2.2 | 2.63 | 1.17 | 2.822 | 4.167 | 1.3 | 2.57 | 2.4 | 2.48 | 3.5 | 3.433 | 1.2 | 1.37 | 2.83 |
| X22 | 9.6 | 10.07 | 9.63 | 9.99 | 11.7 | 12.07 | 10.7 | 10.2 | 10.3 | 11.7 | 11.1 | 9.7 | 11.92 | 11 | 10.9 | 10.4 | 10.8 | 9.37 | 11.02 | 12.37 | 9.5 | 10.8 | 10.6 | 10.7 | 11.7 | 11.63 | 9.4 | 9.57 | 11 |
| X23 | 8.4 | 8.9 | 8.47 | 8.83 | 10.6 | 11 | 9.67 | 9.11 | 9.1 | 10.6 | 10 | 8.533 | 10.9 | 8.9 | 8.91 | 9.367 | 9.75 | 8.2 | 9.956 | 11.33 | 8.3 | 9.67 | 9.43 | 9.48 | 10.7 | 10.6 | 8.23 | 8.4 | 9.87 |
| X24 | 5.8 | 6.267 | 5.83 | 6.19 | 7.9 | 8.267 | 6.9 | 6.41 | 6.47 | 7.87 | 7.3 | 5.9 | 8.117 | 7 | 7.02 | 6.6 | 7.03 | 5.57 | 7.222 | 8.567 | 5.7 | 6.97 | 6.8 | 6.88 | 7.9 | 7.833 | 5.6 | 5.77 | 7.23 |
| X25 | 5.4 | 5.867 | 5.43 | 5.81 | 7.5 | 7.867 | 6.5 | 6.01 | 6.07 | 7.47 | 6.9 | 5.5 | 7.717 | 6.6 | 6.63 | 6.2 | 6.63 | 5.17 | 6.822 | 8.167 | 5.3 | 6.57 | 6.4 | 6.48 | 7.5 | 7.433 | 5.2 | 5.37 | 6.83 |
| X26 | 5.4 | 5.867 | 5.43 | 5.81 | 7.5 | 7.867 | 6.5 | 6.01 | 6.07 | 7.47 | 6.9 | 5.5 | 7.717 | 6.6 | 6.63 | 6.2 | 6.63 | 5.17 | 6.822 | 8.167 | 5.3 | 6.57 | 6.4 | 6.48 | 7.5 | 7.433 | 5.2 | 5.37 | 6.83 |
| X27 | 6.2 | 6.689 | 6.26 | 6.67 | 8.389 | 8.733 | 7.39 | 6.84 | 6.89 | 8.29 | 7.722 | 6.322 | 8.606 | 7.6 | 7.56 | 7.022 | 7.52 | 5.99 | 6.88 | 9.078 | 6.2 | 7.39 | 7.22 | 7.3 | 8.41 | 8.344 | 6.02 | 6.19 | 7.74 |
| X28 | 8.4 | 8.892 | 8.43 | 8.79 | 10.47 | 10.84 | 9.49 | 9.04 | 9.11 | 10.4 | 9.792 | 8.492 | 9.815 | 9.8 | 9.76 | 9.242 | 9.64 | 8.16 | 9.806 | 11.13 | 8.3 | 9.5 | 9.41 | 10.8 | 10.5 | 10.48 | 8.19 | 8.36 | 9.88 |
| X29 | 5.4 | 5.867 | 5.43 | 5.82 | 7.5 | 7.867 | 6.5 | 6.01 | 6.07 | 7.47 | 6.9 | 5.5 | 7.717 | 6.7 | 6.65 | 6.2 | 6.63 | 5.17 | 6.822 | 8.167 | 5.3 | 6.57 | 6.4 | 6.48 | 7.5 | 7.433 | 5.2 | 5.37 | 6.83 |

*Table 6: Total relation matrix*

| | X1 | X2 | X3 | X4 | X5 | X6 | X7 | X8 | X9 | X10 | X11 | X12 | X13 | X14 | X15 | X16 | X17 | X18 | X19 | X20 | X21 | X22 | X23 | X24 | X25 | X26 | X27 | X28 | X29 |
|---|---|---|---|---|---|---|---|---|---|---|---|---|---|---|---|---|---|---|---|---|---|---|---|---|---|---|---|---|---|
| X1 | 0.04 | 0.04 | 0.04 | 0.04 | 0.05 | 0.05 | 0.04 | 0.04 | 0.04 | 0.04 | 0.04 | 0.04 | 0.04 | 0.05 | 0.04 | 0.04 | 0.04 | 0.04 | 0.04 | 0.04 | 0.05 | 0.04 | 0.04 | 0.04 | 0.04 | 0.05 | 0.04 | 0.04 | 0.04 |
| X2 | 0.03 | 0.03 | 0.03 | 0.03 | 0.04 | 0.04 | 0.04 | 0.03 | 0.03 | 0.04 | 0.04 | 0.04 | 0.03 | 0.04 | 0.04 | 0.04 | 0.04 | 0.04 | 0.04 | 0.04 | 0.04 | 0.03 | 0.04 | 0.04 | 0.04 | 0.04 | 0.04 | 0.03 | 0.04 |
| X3 | 0.03 | 0.03 | 0.03 | 0.03 | 0.04 | 0.04 | 0.03 | 0.03 | 0.03 | 0.04 | 0.04 | 0.04 | 0.03 | 0.04 | 0.03 | 0.03 | 0.03 | 0.03 | 0.03 | 0.03 | 0.04 | 0.04 | 0.03 | 0.03 | 0.03 | 0.03 | 0.04 | 0.04 | 0.03 | 0.04 |
| X4 | 0.03 | 0.03 | 0.03 | 0.03 | 0.04 | 0.04 | 0.03 | 0.03 | 0.03 | 0.04 | 0.04 | 0.04 | 0.03 | 0.04 | 0.03 | 0.03 | 0.03 | 0.04 | 0.04 | 0.04 | 0.04 | 0.03 | 0.03 | 0.03 | 0.03 | 0.04 | 0.04 | 0.03 | 0.03 | 0.04 |
| X5 | 0.03 | 0.03 | 0.03 | 0.03 | 0.03 | 0.03 | 0.03 | 0.03 | 0.03 | 0.03 | 0.03 | 0.03 | 0.03 | 0.03 | 0.03 | 0.03 | 0.03 | 0.03 | 0.03 | 0.03 | 0.04 | 0.03 | 0.03 | 0.03 | 0.03 | 0.03 | 0.03 | 0.03 | 0.03 |
| X6 | 0.04 | 0.04 | 0.04 | 0.04 | 0.05 | 0.05 | 0.04 | 0.04 | 0.04 | 0.05 | 0.04 | 0.04 | 0.04 | 0.05 | 0.04 | 0.04 | 0.04 | 0.04 | 0.04 | 0.04 | 0.05 | 0.04 | 0.04 | 0.04 | 0.04 | 0.05 | 0.05 | 0.04 | 0.04 | 0.04 |
| X7 | 0.04 | 0.04 | 0.04 | 0.04 | 0.05 | 0.05 | 0.05 | 0.04 | 0.04 | 0.05 | 0.05 | 0.04 | 0.05 | 0.05 | 0.05 | 0.04 | 0.05 | 0.04 | 0.05 | 0.05 | 0.04 | 0.05 | 0.05 | 0.05 | 0.05 | 0.05 | 0.04 | 0.04 | 0.05 |
| X8 | 0.05 | 0.05 | 0.05 | 0.05 | 0.06 | 0.06 | 0.06 | 0.05 | 0.05 | 0.06 | 0.06 | 0.05 | 0.06 | 0.06 | 0.06 | 0.05 | 0.06 | 0.05 | 0.06 | 0.06 | 0.05 | 0.06 | 0.06 | 0.06 | 0.06 | 0.06 | 0.05 | 0.05 | 0.06 |
| X9 | 0.05 | 0.05 | 0.05 | 0.05 | 0.06 | 0.06 | 0.05 | 0.05 | 0.05 | 0.06 | 0.06 | 0.05 | 0.06 | 0.06 | 0.06 | 0.05 | 0.06 | 0.05 | 0.06 | 0.06 | 0.05 | 0.06 | 0.06 | 0.06 | 0.06 | 0.06 | 0.05 | 0.05 | 0.06 |
| X10 | 0.04 | 0.04 | 0.04 | 0.04 | 0.04 | 0.05 | 0.04 | 0.04 | 0.04 | 0.04 | 0.04 | 0.04 | 0.04 | 0.04 | 0.04 | 0.04 | 0.04 | 0.04 | 0.03 | 0.04 | 0.05 | 0.04 | 0.04 | 0.04 | 0.04 | 0.04 | 0.03 | 0.04 | 0.04 |
| X11 | 0.04 | 0.04 | 0.04 | 0.04 | 0.05 | 0.05 | 0.04 | 0.04 | 0.04 | 0.04 | 0.05 | 0.04 | 0.04 | 0.05 | 0.04 | 0.04 | 0.04 | 0.04 | 0.04 | 0.04 | 0.05 | 0.04 | 0.04 | 0.04 | 0.04 | 0.05 | 0.05 | 0.04 | 0.04 | 0.04 |
| X12 | 0.03 | 0.04 | 0.03 | 0.04 | 0.04 | 0.04 | 0.04 | 0.04 | 0.04 | 0.04 | 0.04 | 0.04 | 0.03 | 0.04 | 0.04 | 0.04 | 0.04 | 0.03 | 0.04 | 0.04 | 0.03 | 0.04 | 0.04 | 0.04 | 0.04 | 0.04 | 0.04 | 0.03 | 0.03 | 0.04 |
| X13 | 0.03 | 0.04 | 0.03 | 0.04 | 0.04 | 0.04 | 0.04 | 0.04 | 0.04 | 0.04 | 0.04 | 0.04 | 0.03 | 0.04 | 0.04 | 0.04 | 0.04 | 0.04 | 0.03 | 0.04 | 0.04 | 0.04 | 0.04 | 0.04 | 0.04 | 0.04 | 0.04 | 0.03 | 0.03 | 0.04 |
| X14 | 0.04 | 0.04 | 0.04 | 0.04 | 0.04 | 0.05 | 0.04 | 0.04 | 0.04 | 0.04 | 0.04 | 0.04 | 0.04 | 0.04 | 0.04 | 0.04 | 0.04 | 0.04 | 0.03 | 0.04 | 0.05 | 0.04 | 0.04 | 0.04 | 0.04 | 0.04 | 0.04 | 0.03 | 0.04 | 0.04 |
| X15 | 0.04 | 0.04 | 0.04 | 0.04 | 0.05 | 0.05 | 0.04 | 0.04 | 0.04 | 0.04 | 0.05 | 0.04 | 0.04 | 0.05 | 0.04 | 0.04 | 0.04 | 0.04 | 0.04 | 0.04 | 0.05 | 0.04 | 0.04 | 0.04 | 0.04 | 0.05 | 0.05 | 0.04 | 0.04 | 0.04 |
| X16 | 0.07 | 0.07 | 0.07 | 0.07 | 0.08 | 0.08 | 0.07 | 0.07 | 0.07 | 0.08 | 0.08 | 0.07 | 0.08 | 0.07 | 0.07 | 0.07 | 0.07 | 0.07 | 0.07 | 0.07 | 0.08 | 0.07 | 0.07 | 0.07 | 0.07 | 0.08 | 0.08 | 0.07 | 0.07 | 0.07 |
| X17 | 0.04 | 0.04 | 0.04 | 0.04 | 0.05 | 0.05 | 0.05 | 0.04 | 0.04 | 0.04 | 0.05 | 0.04 | 0.05 | 0.04 | 0.04 | 0.04 | 0.04 | 0.04 | 0.04 | 0.05 | 0.04 | 0.05 | 0.05 | 0.04 | 0.05 | 0.05 | 0.05 | 0.04 | 0.04 | 0.05 |
| X18 | 0.04 | 0.04 | 0.04 | 0.04 | 0.04 | 0.04 | 0.04 | 0.04 | 0.04 | 0.04 | 0.04 | 0.04 | 0.04 | 0.05 | 0.04 | 0.04 | 0.04 | 0.04 | 0.04 | 0.04 | 0.03 | 0.04 | 0.05 | 0.04 | 0.04 | 0.04 | 0.04 | 0.04 | 0.04 | 0.04 |
| X19 | 0.04 | 0.04 | 0.04 | 0.04 | 0.04 | 0.04 | 0.04 | 0.04 | 0.04 | 0.04 | 0.04 | 0.04 | 0.04 | 0.05 | 0.04 | 0.04 | 0.04 | 0.04 | 0.04 | 0.04 | 0.04 | 0.04 | 0.04 | 0.04 | 0.04 | 0.04 | 0.04 | 0.04 | 0.04 | 0.04 |
| X20 | 0.03 | 0.04 | 0.04 | 0.04 | 0.04 | 0.04 | 0.04 | 0.04 | 0.04 | 0.04 | 0.04 | 0.04 | 0.04 | 0.04 | 0.04 | 0.04 | 0.04 | 0.04 | 0.04 | 0.04 | 0.03 | 0.04 | 0.04 | 0.04 | 0.04 | 0.04 | 0.04 | 0.03 | 0.03 | 0.04 |
| X21 | 0.01 | 0.01 | 0.01 | 0.01 | 0.01 | 0.01 | 0.01 | 0.01 | 0.01 | 0.01 | 0.01 | 0.01 | 0.01 | 0.01 | 0.01 | 0.01 | 0.01 | 0.01 | 0.01 | 0.01 | 0.01 | 0.01 | 0.01 | 0.01 | 0.01 | 0.01 | 0.01 | 0.01 | 0.01 |
| X22 | 0.04 | 0.04 | 0.04 | 0.04 | 0.04 | 0.05 | 0.04 | 0.04 | 0.04 | 0.04 | 0.04 | 0.04 | 0.04 | 0.04 | 0.04 | 0.04 | 0.04 | 0.04 | 0.04 | 0.04 | 0.05 | 0.04 | 0.04 | 0.04 | 0.04 | 0.04 | 0.04 | 0.04 | 0.04 | 0.04 |
| X23 | 0.03 | 0.03 | 0.03 | 0.03 | 0.04 | 0.04 | 0.04 | 0.03 | 0.03 | 0.04 | 0.04 | 0.04 | 0.03 | 0.04 | 0.04 | 0.04 | 0.04 | 0.04 | 0.04 | 0.03 | 0.04 | 0.04 | 0.04 | 0.04 | 0.04 | 0.04 | 0.04 | 0.03 | 0.03 | 0.04 |
| X24 | 0.02 | 0.02 | 0.02 | 0.02 | 0.03 | 0.03 | 0.03 | 0.02 | 0.02 | 0.03 | 0.03 | 0.02 | 0.03 | 0.03 | 0.03 | 0.03 | 0.03 | 0.02 | 0.03 | 0.03 | 0.02 | 0.03 | 0.03 | 0.03 | 0.03 | 0.03 | 0.03 | 0.02 | 0.02 | 0.03 |
| X25 | 0.02 | 0.02 | 0.02 | 0.02 | 0.03 | 0.03 | 0.02 | 0.02 | 0.02 | 0.03 | 0.03 | 0.02 | 0.03 | 0.03 | 0.03 | 0.02 | 0.03 | 0.02 | 0.03 | 0.03 | 0.02 | 0.02 | 0.02 | 0.02 | 0.02 | 0.03 | 0.03 | 0.02 | 0.02 | 0.03 |
| X26 | 0.02 | 0.02 | 0.02 | 0.02 | 0.03 | 0.03 | 0.02 | 0.02 | 0.02 | 0.03 | 0.03 | 0.02 | 0.03 | 0.03 | 0.03 | 0.02 | 0.03 | 0.02 | 0.03 | 0.03 | 0.02 | 0.02 | 0.02 | 0.02 | 0.03 | 0.03 | 0.02 | 0.02 | 0.03 |
| X27 | 0.02 | 0.03 | 0.02 | 0.03 | 0.03 | 0.03 | 0.03 | 0.03 | 0.03 | 0.03 | 0.03 | 0.03 | 0.02 | 0.03 | 0.03 | 0.03 | 0.03 | 0.02 | 0.03 | 0.03 | 0.02 | 0.03 | 0.03 | 0.03 | 0.03 | 0.03 | 0.03 | 0.02 | 0.02 | 0.03 |
| X28 | 0.03 | 0.03 | 0.03 | 0.03 | 0.04 | 0.04 | 0.04 | 0.03 | 0.03 | 0.04 | 0.04 | 0.03 | 0.04 | 0.04 | 0.04 | 0.03 | 0.04 | 0.03 | 0.04 | 0.04 | 0.03 | 0.04 | 0.04 | 0.04 | 0.04 | 0.04 | 0.04 | 0.03 | 0.03 | 0.04 |
| X29 | 0.02 | 0.02 | 0.02 | 0.02 | 0.03 | 0.03 | 0.02 | 0.02 | 0.02 | 0.03 | 0.03 | 0.02 | 0.03 | 0.03 | 0.03 | 0.02 | 0.03 | 0.02 | 0.03 | 0.03 | 0.02 | 0.03 | 0.03 | 0.02 | 0.02 | 0.03 | 0.03 | 0.02 | 0.02 | 0.03 |

The elements whose net effect (ri-ci) is a positive value, constitute the cause group. The positivness of the net effect of each factor means that its influential degree is more than its influenced degree ( the effect it receives from othe system factors). Elements of this group include, innovation itself (X16), signal quality (X8), transmission speed (X9), security and privacy (X7),handset price (X1), time (X18), advertising (X17), handset adaptibility (X12), uncertainty (X15), perceived ease of use (X11), age (X28), professional realtionship (X22), operator trust (X14), social environment (X19) and transmission cost (X15).

*Table 7:Score of each factor*

| | Factors | Ri+Cj | Ri-Cj | Cj | Ri |
|---|---|---|---|---|---|
| X1 | Phone price (hand set) | 2.174 | 0.202 | 0.986 | 1.188 |
| X2 | Transmission cost (call cost) | 2.083 | 0.007 | 1.038 | 1.045 |
| X3 | Network coverage | 1.972 | -0.01 | 0.99 | 0.982 |
| X4 | Communication channels (press and media | 2.032 | -0.03 | 1.032 | 1 |
| X5 | Friends and relatives' recommendations | 2.087 | -0.35 | 1.219 | 0.868 |
| X6 | number of users | 2.482 | -0.04 | 1.26 | 1.221 |
| X7 | Security and privacy | 2.442 | 0.222 | 1.11 | 1.332 |
| X8 | Signal quality | 2.675 | 0.564 | 1.056 | 1.619 |
| X9 | Transmission speed | 2.68 | 0.56 | 1.06 | 1.62 |
| X10 | Usefulness (perceived usefulness) | 2.363 | -0.07 | 1.215 | 1.148 |
| X11 | Perceived ease of use | 2.397 | 0.092 | 1.152 | 1.244 |
| X12 | Device (phone)adaptability | 2.109 | 0.115 | 0.997 | 1.112 |
| X13 | Company's reputation (validity) | 2.353 | -0.13 | 1.242 | 1.111 |
| X14 | trust | 2.266 | 0.045 | 1.111 | 1.156 |
| X15 | uncertainty | 2.333 | 0.111 | 1.111 | 1.222 |
| X16 | -innovation itself (technical superiority ) | 3.18 | 1.026 | 1.077 | 2.103 |
| X17 | advertisement | 2.436 | 0.188 | 1.124 | 1.312 |
| X18 | time | 2.119 | 0.198 | 0.96 | 1.158 |
| X19 | Social environment | 2.32 | 0.035 | 1.142 | 1.177 |
| X20 | attitude | 2.428 | -0.16 | 1.294 | 1.133 |
| X21 | Perceived behavioral control | 1.247 | -0.7 | 0.975 | 0.272 |
| X22 | Professional relationship | 2.293 | 0.062 | 1.116 | 1.177 |
| X23 | Result demonstrability | 2.144 | -0.05 | 1.097 | 1.048 |
| X24 | Long term outcomes | 1.866 | -0.35 | 1.109 | 0.757 |

| X25 | Performance expectance | 1.934 | -0.51 | 1.221 | 0.713 |
| X26 | Effort expectance | 1.927 | -0.5 | 1.213 | 0.713 |
| X27 | gender | 1.769 | -0.16 | 0.964 | 0.805 |
| X28 | age | 2.028 | 0.063 | 0.982 | 1.046 |
| X29 | experience | 1.858 | -0.43 | 1.145 | 0.714 |

The elements that comprise the negative net effect are the effect group. This group includes the factors of network coverage (X3), communication channels (X4), number of users (X6), result demonestrability (X2), perceived usefulness (X10), company's reputation (validity) (X13), gender (X27), attitude (X20), friends and relatives' recommendations (X5), long-term outcomes (X24), effort expectance (X26), performance expectance (X25) and Perceived Controlling Control (X21).There is a lot of information in Figure 4 that can be used for decision making. Identificatio of critical factorsaffecting the adoption of 4G technology in the Iranian society is performed using information from
Table 7and Figure 5

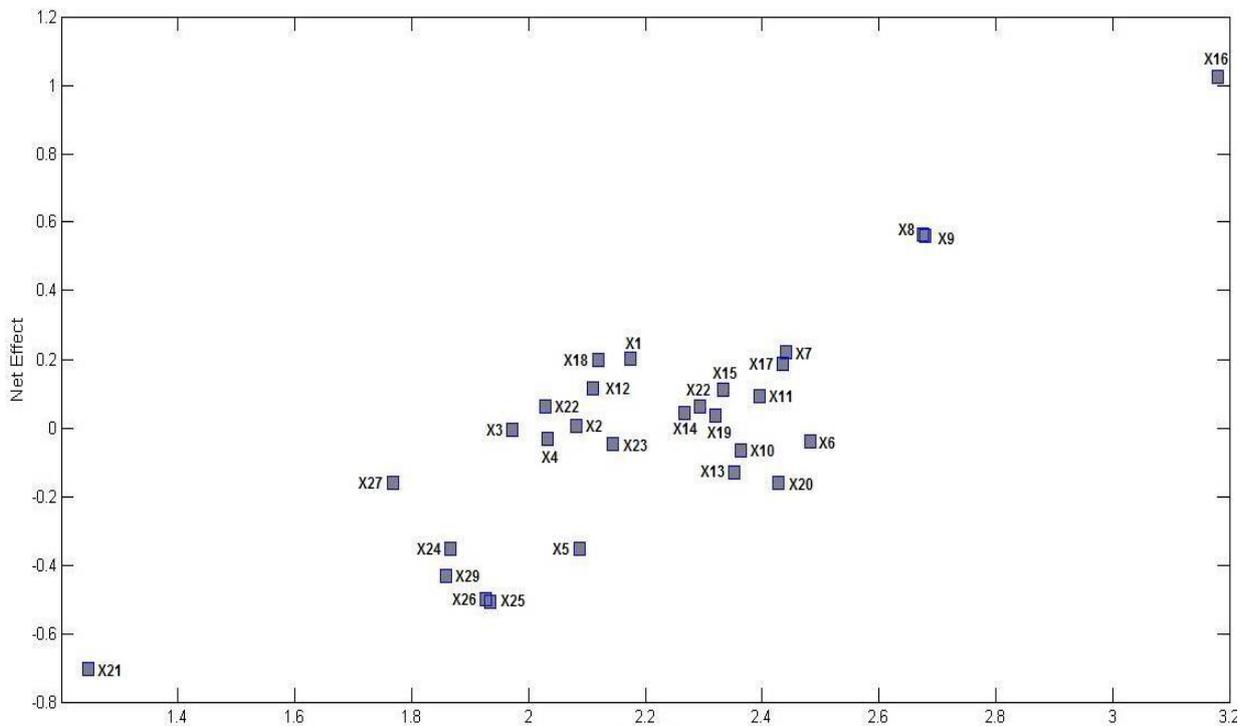

*Figure 5:Cause and effect relation diagram*

*Discussion*

The "innovation itself" factor (6X1) is the first factor affecting the decision of Iranian users to adopt or not to adopt fourth generation technology. This factor has the highest net effect, i.e. the influential degree of this factor on the whole system is in turn more than its influenced degree. In addition, the net effect of this factor with its next factor i.e the signal quality (X8) is very high (0.462), which indicates that this factor has a very significant effect in encouraging the Iranian users to use LTE technology instead of UMTS / HSPA technologies. Rogers defines this factor as a general category, which is basically referred to as an idea, a mode of action, or anything new or novel. [39] Therefore, this factor can be considered equivalent to the "newly perceived". That is, users should consider a technology more innovative than competing technologies. However, this concept may be synonymous with concepts such as technical superiority but it should be considered that innovation itself or newly perceived is something distinct from technical superiority, so that a " newly perceived " refers to a level in which an individual knows something new, but the concept of technical superiority is a composite index of the superior characteristics of a technology, such as the transmission speed (X7)and signal quality (9X),for this case, depending on the user's knowledge, they can be varied. For example, for a user who has no engineering knowledge about mobile networks, the superiority of LTE fourth generation mobile networks in comparison with UMTS third generation may be the higher speed of LTE in comparison with previous technology. While a person specializing in mobile networks may consider a variety of differences, as an example, one can see one of the technological superiorities of LTE mobile networks in the fact that the fourth generation uses the OFDM method to allocate a frequency spectrum among users, while the UMTS / HSPA generation uses the WCDMA method to do this.

"Signal quality and network coverage (X9)" is the second most influential factor in the diffusion of fourth-generation technology in the Iranian community. According to the 1395 census, the population of Iran in 1995 was over 80 million, and due to the fact that the land area of this country is equivalent to 162,886 square kilometers and its population density is 50, the vast size of the country with its population is implied, which confirms the role of important factors such as "signal quality and network coverage" (X9).

"Transmission speed (X7)" is also the third factor affecting the decision of Iranian users to accept or reject the fourth generation technology. Principally, in Iran's community, the main difference between the third and fourth generations is related to its speed and other technical differences are not considered. With the increasing development of OTT content and the strong proliferation of instant messenger services such as Telegram with a penetration rate of 78% and Instagram with a penetration rate of 54% in the Iranian community, the role and importance of speed of mobile data provision has become one of the most basic demands of Iranian citizens from a mobile operator, and the results of this study illustrate this point. In the third-generation UMTS technology, the maximum link speed of rising-up is 384Kbps and the maximum falling-down speed is 128Kbps, while the falling-down and rising-up speed has improved dramatically in the fourth generation.In the fourth-generation LTE, the maximum rising-up speed is 10Mbps and the maximum rising-up speed is up to 500Mbps.

"The security and privacy of users (X8)" is the fourth determining factor in the diffusion of fourth-generation technology in Iranian community. Internet privacy includes the privacy of internet users for the storage of information and their personal use, providing third parties or displaying in internet space. In addition to information, this privacy is also related to a person who is in charge of space, i.e the place a person is associated with. At the end of spring 2011, smartphones became common in. Iran, which has led to more people using OTT technologies. So that until the end of spring 2017, over 36% of people use what's app, 54% of people use Instagram and 78% of people use Telegram. OTT technologies provide users with various communication channels through which they can produce different contents. These contents can be presented in the form of user comments, which can affect the demands of other users. This affect can be varied from any OTT technology such as a telegram to any other technology such as facebook. By the end of the spring 2017, about 55% of the Iranian community use the high-speed Internet, 10 million use fixed broadband and 34 million use mobile broadband, that 77% of its users use mobile broadband. The majority of the population using the Internet is the active population of community, aged between 15 and 65. However, statistics show that in Iran online social networks have become an important part of online activities on the web. These networks with an unlimited physical distance provide web users, new tools for communicating, interacting, and socializing. However, although these networks provide the ability of data sharing and, in an instant way, enable communication between users, if not properly used, poses many challenges for the privacy of individuals.

"Phone price (X1)" is the fifth determining factor in the diffusion of fourth-generation technology in the Iranian community. Every mobile handset cannot support the fourth-generation technology, (which is definitely a mobile LTE).handsets that support this technology are at a higher price compared to their brandedhandsets, which only support previous generations (such as GSM and UMTS). Therefore, thehandset price factor, depending on the difference in people's incomes, can affect the diffusion of advanced mobile generations.

"Time (X18)" is the sixth determining factor in the diffusion of fourth-generation technology in. Iranian society. Diffusion speed and rate of technology has varying degrees over time. Rogers, in his bell curve, points out that technology adoption will be different over time. So that, as time goes by, different categories of people adopt innovation. He showed that when a new technology is introduced, only 2.5% of the people adopt it. He call such people as innovators. Over time 13.5% of people will adopt the technology. Rogers calls these people as "early adopters". Next, 34% of the people will adopt technology, Rogers calls these people as an "early majority." Over time, technology will penetrate another part of the population, including 34% of the population, which Rogers calls the "late majority". As technology penetrates more over time, the remaining 16 % of society, whom Rogers calls "laggard", will ultimately adopt the innovation. One point to note is that this curve differs from one technology to another technology. This situation can be seen in mobile and fixed broadband technologies, as mobile broadband technologies have so far been roughly adopted by 50% of the population, covering the majority of the early majority. While fixed broadband technologies have so far only been adopted by 11.9% of the community, i.e. they still have not fully covered the early adopters. By the end of the spring 2017, the fourth generation technology has been adopted by at most 6% of the Iranian population. This situation indicates that the technology, in regards with technology adotion curve has already been removed from the "early adopters".

"Advertising (X17)" is the seventh determining factor in the diffusion of fourth-generation technology in Iranian society. Mass media play an important role in familiarizing people with new facts, so that the familiarity of individuals with superiority of a technology can influence their decision to adopt it. At the beginning of this study, it was believed that advertising was among the first three factors influencing the diffusion of fourth-generation technology in Iranian society, but after obtaining the results of the research, it became clear that advertising was the ninth factor. The reason for this result is more likely due to the heterogeneity of Iranian market and its cultural differences (its specific cultural features) from other countries whose advertising template advertisers make use of.Graig and Douglas, Maxwell and Rawwas, in their studies, referred to the role and importance of cultural differences in the effectiveness of advertising. They believe that advertising should be designed based on the needs and cultural differences of societies in order to have more impact, and one cannot create a standard advertising model for all communities.

"Device adaptability (X12)" is the eighth factor affecting the decision of. Iranian users to accept or reject fourth-generation technology. Due to technological differences, any mobile handset cannot support fourth-generation technology, hence thehandset's adaptability with network operating system has an importance that should be taken into account in network diffusion. In order to penetarate the market, two mobile operators presented a free SIM card (USIM), but contrary to the expectation, they did not lead to third generation users to use the fourth generation of mobile handsets, one of the main obstacles of which might be the lack of adaptability and lack of support for subscribers of the fourth generation networks.

"Uncertainty (X15)" is the ninth factor affecting the decision of Iranian users to accept or reject the fourth generation technology. Uncertainty in the advantages of fourth-generation technology can slow down its adoption by subscribers. The fourth-generation technology has superiority over third-generation technology from different aspects such as latency and data speed. Accordingly, perhaps the most important function of advertising factor (X17) is to increase the level of certainty of third-generation subscribers in the fourth-generation technological superiority.

The perceived ease of use (X11) is the tenth factor affecting the adoption of the fourth generation technology in Iranian society, which referrs to the ease of use that a person considers a technology use. The more easy to use an individual considers the technology use, i.e its use requires less effort, the more tendency to use it will occur.

"Age (X28)" is the eleventhth factor in the diffusion of fourth-generation technology in Iranian society. Generally, age groups of 15- 29 and 30-64 are more inclined to adopt modern technologies than age groups under the age of 15 or over 65. The first two age groups, which include 69.9% of I.R. Iran's population, are the active population of the country, and the size of this population is a great potential for adopting new technologies.

"Occupation relation(X22)" is the twelfth factor in the diffusion of fourth-generation technology in Iranian society, the more people's occupation requires using technological advatages, such as lower latency, data speed, upload and download, Of course, the more tendency people have to use a suitable technology. By the end of 1395, 4,348,386 people have been studying at the university, for whom mobile broadband especially fourth generation can have a lot of attraction.

"Operator trust (X14)", is the thirteenth factor in the diffusion of fourth-generation technology in Iranian society. Trust is basically referred to the community's perception of the operators. Mobile operators that have not created a good experience for the community, may not be able to attract an acceptable portion of the community even if they develop the best mobile network technology.

"Social environment (X20)" is the fourteenth factor affecting the decision of Iranian users to accept or reject the fourth generation technology. The social environment is the sum of all social factors such as age, gender, beliefs, cultural norms, expectations, class divisions, and so on that has an undeniable effect on the decision of community's people. The social environment is a macro variable that includes all subscribers and non-subscribers of fourth-generation technology in I.R. Iran community. The larger the number of a network users, the greater benefit the next subscribers of that network will have. The voice and SMS cost for intra-network subscribers, is different from that for the inter-network subscriber. For example, subscribers who use I.R. Irancell operator pay for the voice and text (intranet) communication less , but if an I.R. Irancell subscriber calls an MCI subscriber should pay more inter-network cost. Network status in terms of voice and SMS cost can be of benefit to subscribers of each operator (independent of the type of mobile technology they use). But this can be two-way for mobile data. On one hand, the subscribers of each operator, using mobile data provided by that operator, are connected to a single network (Internet) through which they can communicate with each other using different OTT platforms (audio, text and video). In this case, the greater the number of users of these platforms, the greater benefit the next users will have from these platforms, even if their users are connected to the Internet with different mobile technologies. In another case, the dedicated data of any mobile technology can create a dedicated network for its users. As an example, in countries that services such as voice on LTE platform and video on LTE platform have been launched, LTE technology users will be able to take advantage of the positive effects of these services, provided theirhandsets support these services. Second and third-generation networks cannot provide voice and video services on their data bases, because they are developed based on circuit switching meyhodology; however in fourth-generation networks that have been developed based on the "small packet switching" methodology, there is the possibility of developing audio and video services on their data bases, and they themselves can have a positive network effects for their users.

"Transmission cost (X2)" is the fifteenth determining factor in the fourth-generation technology in Iranian society. In I.R. Iran, voice and SMS services are regulated by Communication Regulatory Authority (CRA), so the pricing of this service is not generally under the authority mobile operators. Nevertheless, operators can independently price their mobile internet value-added services. Mobile data pricing strategy is very important for operators in I.R. Iran, so that by adopting appropriate prices and suitable volumes, mobile data can penetrate different parts of the market.

"Attitude (X20)" is the 16th factor that affects the decision of Iranian users to accept or reject the fourth generation technology. Each user's attitude refers to the angle he looks at technology. This factor is a composite variable and more than any other factor is influenced by other factors. Its influenced score is 1.294. The main reason that this factor's net effect score is negative (0.161) is that its influential degree (1.133) is less that its influenced degree. With the knowledge of the above-mentioned fifteen key factors, telecom managers need to plan in a way that can direct their operator's subscribers' attitude in a way that those who do not use the fourth generation tend to be more inclined toward it. All critical factorscan be intuitively categorized into four groups. These groups include the "group of mobile handset & operators-related factors ", the "group of Subscribers-related biological factors ", " group of subscribers-related perceptual factors " and "group of subscribers-related contextual factors ". Group of mobile handset & operators-related factors include factors such as signaling quality (X8), transmission speed (X9), mobile handset price (X1), device adaptability(X12) and data transmission cost (X2). The group of subscribers-related biological factors include the factor of age (X28). The group of subscribers-related perceptual factors includes factors such as innovation-newly perceived (X16), security and privacy (X7), uncertainty about technological advatages (X15), perceived easeof use (X11), occupation relation (X22), trust to operators (X14), and attitude (X20). The group of subscribers-related contextual factors includes factors such as time (X18), advertising (X17) and social environment (X19). In order to facilitate the diffusion of fourth-generation LTE technology in I.R. Iran, all of these four groups should be considered by the telecom industry planners

V. CONCLUSION AND DIRECTIONS OF FUTURE STUDIES

In spite of the large investments of the three main operators of I.R. Iran in the development of the fourth generation network, as well as the large penetration of mobile broadband in this country (about 42.5% at the end of spring 2017), the LTE mobile penetration rate in I.R. Iran has been estimated to be 0.06, indicating that the main mobile broadband subscribers in I.R. Iran, about 29 million, still use third generation UMTS / HSPA networks. If operators fail to encourage their subscribers to use LTE's fourth-generation network, there will be plenty of losses. To overcome this challenge, operator planners have to identify the effective factors leading to facilitation of fourth generation network adoption in the community. A detailed study of literature showed that several factors are involved in adopting an innovation. In this study, 29 factors were identified. However, due to the fact that organizations' resources are limmited , they can not simultaneously consider all of these factors in their planning. Therefore, it was very important to know critical success factors. In this research, all the factors affecting technology adoption were evaluated through the fuzzy DEMATEL method. Using the results of this method, all factors were divided into cause and effect groups. In addition, a cause-effect diagram was also extracted. In the end, sixteen factors were considered as critical factorsin the adoption of technology in Iranian society. These critical factorswere intuitively divided into four groups. To accelerate the diffusion of fourth-generation networks in I.R. Iran, operator planners should pay particular attention to these factors.

However, the outputs of this method are at macro level. They only make planners aware of the critical factors affecting adoption. That is, how each factor independently influences the adoption rate of the fourth generation technology. To understand how these factors influence the adoption of the fourth-generation technology, we need to make a smaller-level network analysis. By the help of this approach, it is possible to simulate the network in which these factors interact and ultimately change the level of technology adoption. A network analysis that has a smaller level enables managers to better understand how each factor change influence the entire network and ultimately the extent to which the technology is adopted. Through this approach,scenarios of -what if- can be developed. So doing a network analysis of these factors and extracting the most appropriate scenarios of -what if- can be a topic for future research.

In addition, the clustering of 16 key success factors can be another topic for future research. The outputs of the proposed method are presented as 16 separate factors, which are intuitively divided into four groups: "group of mobile handset & operators-related factors ", "group of subscribers-related biological factors ", " group of subscribers-related perceptual factors " and "group of subscribers-related contextual factors ". If a structural clustering of these factors takes place, a more systematic understanding of these factors occurs. In this case, the factors that are close to each other are in the same cluster and managers can better decide on the amount of resource allocation to them.